\documentclass[5p]{elsarticle}
\usepackage{lineno}
\modulolinenumbers[5]
\usepackage{graphicx}

\usepackage{xr}
\usepackage[hidelinks]{hyperref}
\makeatletter
\newcommand*{\addFileDependency}[1]{
  \typeout{(#1)}
  \@addtofilelist{#1}
  \IfFileExists{#1}{}{\typeout{No file #1.}}
}
\makeatother

\usepackage{lmodern}
\usepackage{multirow}
\usepackage{epsfig,mathrsfs,color,latexsym,subfigure,marginnote,textcomp,gensymb}
\usepackage{amsmath}
\usepackage{gensymb}
\usepackage[numbers]{natbib}
\biboptions{sort&compress}
\journal{Materials Today Commnunications}

\begin{document}

\begin{frontmatter}

\title{Microstructure-property prediction of a Ni-based superalloy: A combined phase-field and finite element modelling approach}
\author[a]{Rupesh Chafle \corref{cor2}}
\author[a]{Vishal Panwar \corref{cor2}}
\author[b]{Kaushik Das}
\author[a]{Somnath Bhowmick}
\author[a]{Rajdip Mukherjee\corref{cor1}} 
\address[a]{Department of Materials Science and Engineering, Indian Institute of Technology Kanpur, Kanpur 208016, India}
\address[b]{Department of Metallurgy and Materials Engineering, Indian Institute of Engineering Science and Technology Shibpur, Howrah 711103, India}
\cortext[cor1]{Corresponding author \\E-mail address:rajdipm@iitk.ac.in (R.Mukherjee)}
\cortext[cor2]{These authors contributed equally to this work}


%
%

\begin{abstract}
 Multiscale modelling is a new paradigm that has emerged in recent times to study the well-known problem of the process-structure-property relationship in the area of materials science and engineering. For obtaining the desired performance for materials of strategic importance, such as superalloys, it is essential to bridge different length and time scales in order to navigate the entire design space.  In the present study, we develop a physics-based model for a Ni-based superalloy where the microstructures simulated using a phase-field model serve as input to finite-element computations. We examine the alloy's microstructure evolution and effective elastic properties quantitatively via phase-field and finite element methods integrated with CALPHAD database, by varying composition and aging temperature. The phase-field simulations provide us with an insight into the different regimes of microstructure evolution. The finite element analysis uncovers the relation of effective elastic properties with several system parameters.

\end{abstract}


\end{frontmatter}

\section{Introduction}
Multiscale materials modelling is a problem-solving tool that entails characterising system behaviour at a specific temporal and spatial scale using data from lower length scales, provided that the physics of the problem significantly affects the material's behaviour at the higher length scale. In a multiscale study of the spatio-temporal evolution of microstructure, lower length scale description involves information about electrons, atoms, molecules, and their assemblies, which gives rise to  individual phases and grains. Traditional modelling approaches are phenomenological and use empirical data to depict system behaviour on a continuum scale. On the other hand, Multiscale modelling attempts to map the system's continuum behaviour in terms of the positions and velocities of its constituents, such as atoms and molecules. 

Several approximation theories are used to develop multiscale theory, including the Cauchy–Born rule~\cite{Ericksen2008199}, the Voigt assumption~\cite{Voigt1889573}, the Reuss constant stress hypothesis~\cite{Reuss192949} etc. Upscaling and resolved-scale approaches are the two primary categories of multiscale modelling methods~\cite{Fish2021774}. Lower length scale response is estimated, and its average effect is conveyed to macroscale in upscaling methods; however, in resolved-scale approaches, the problem is solved at lower length scale and macroscale concurrently in distinct domains of the problem. Upscaling can be divided into three types: math-based, physics-based, and data-driven, as well as the direction of information flow across scales (one way or two way). The reduced scale approach has subcategories on the basis of the flow of information transfer. If it is done through the interface between domains of different scales then it is known as Domain decomposition, or if it is done through an interscale transfer operator, then it is known as the Multigrid method. In the present study, we use a physics-based upscaling method for effective property prediction from the simulated microstructures of Ni-Al alloys.

Commercial Ni-Al alloys with two-phase $\gamma$-$\gamma'$ structures have been widely accepted as structural materials in various industrial sectors, e.g. forging dies, helium reactors, heat exchangers and turbine blades~\cite{Jozwik20152537,Yu2016107748}. Nickel-based superalloys can be considered excellent candidates for industrial applications like gas turbines, aircraft, space industry, chemical process plants etc.due to their outstanding load-bearing properties even at 0.8$T_m$. They are the most fascinating superalloys because they balance economics and performance~\cite{Asato2011}. These alloys have high phase stability of the $fcc$ nickel matrix and can be strengthened by solid solution strengthening as well as precipitation hardening. The Ni-based superalloys have yield strength and ultimate tensile strength in the range of 900 to 1300 MPa and 1200 to 1600 MPa respectively at 298K~\cite{reed_2006}. In the present study, we restrict ourselves to only the $\gamma$ and $\gamma'$ phases in a binary Ni-Al system. 

\begin{figure}[ht]
\centering
  \includegraphics[clip, trim=0cm 0cm 10cm 0cm, width = 0.8\columnwidth]{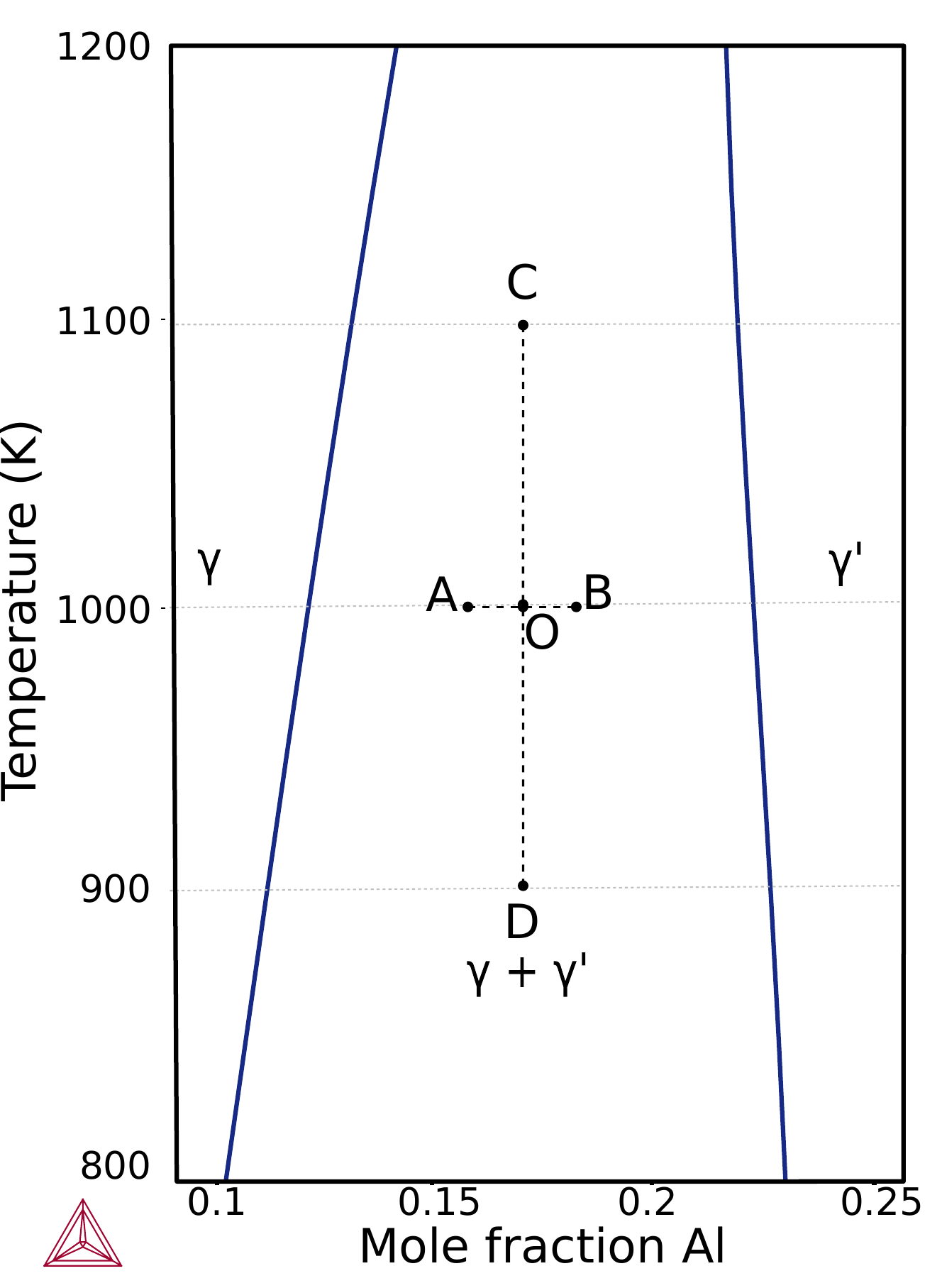}
\caption{A part of the Ni-Al phase diagram obtained from CALPHAD database showing the $\gamma$, $\gamma$ + $\gamma'$ and $\gamma'$ phase-fields~\cite{ANDERSSON2002273}. Points A, B, C, D, and O represent the alloys considered in this work.}
\label{NiAl_PD}
\end{figure}

Fig.~\ref{NiAl_PD} shows a part of the Ni-Al phase diagram with $\gamma$, $\gamma$+$\gamma'$ and $\gamma'$ phase fields~\cite{Tien1971a,Tien1971b, Miyazaki1979,Fratzl1999}. The Ni-Al alloys with $\gamma$-$\gamma'$ microstructure are important structural materials as they have high temperature creep resistance. The $\gamma$ phase is a continuous matrix phase and has a face-centred cubic structure. 
$\gamma'$  is an ordered phase of Ni$_3$Al ($L1_{2}$ structure) where Al atoms occupy the corner positions, and Ni atoms are placed at the centres of the faces. This way, each Ni atom has four Al atoms, and eight Ni atoms as nearest neighbours and each Al atom is coordinated by twelve Ni atoms. The $\gamma'$ phase particles are cuboidal coherent particles homogeneously dispersed in the matrix of disordered $\gamma$ leading to the strengthening of the matrix. 
The $\gamma'$ precipitates align along the elastically soft $<$100$>$ direction. Interaction of dislocations and precipitates contributes to the strengthening in these alloys; the dislocations either have to cut the precipitate or by-pass them (Orowan looping) \cite{PhysRevLett.70.2305, Wang2004933, MATAN19992031}. \textcolor{black}{The $\gamma$-$\gamma'$ phases are compatible with each other with 0.1-1.0\% mismatch in their lattice parameters and therefore have low interfacial energy ($\sim$10 $mJ m^{-2}$).} Minimization of total interfacial energy drives the process of precipitate coarsening. Hence a coherent interface ensures the stability of the microstructure (useful for high-temperature applications).

The phase field approach, combined with CALPHAD and Density functional theory (DFT), can model the microstructural evolution. The Density Functional Theory is a method to approximately solve the Schr\"{o}dinger equation of a many-electron system~\cite{Kohn1965A1133}. It is predominantly used to predict structural, electronic, magnetic, and optical properties in materials, which are further utilised for microstructural evolution and material behaviour predictions. CALPHAD integration provides the appropriate free energy functional~\cite{kaufman1970computer}. The diffuse-interface approach incorporated here has been used to study a wide range of processes of practical importance~\cite{Chen2002113, Koyama2009891}. Some prominent examples are in solidification~\cite{Karma2001, Tang201619773}, martensitic transformation~\cite{Artemev20011165}, domain formation and growth
 ~\cite{Wang2004749,chafle2022}, Ostwald ripening~\cite{Fan20021895,chakrabarti2019,chafle2020}, recrystallization~\cite{Chen2015829}, dislocation dynamics~\cite{Jin2001607}, grain boundary diffusion~\cite{LVOV2023106209}, spinodal decomposition~\cite{CAHN1961795,CAHN1962179,TILLER1970225,CAHN1971151,chafle2019,GUO2023105811}, microstructure evolution in thin films~\cite{miral2021jap,mukherjee2022}, and crack propagation~\cite{Eastgate2002, Miehe20102765}.

The effective elastic properties are not only influenced by the properties of the phases but are also sensitive to the microstructure. Thus, predicting the effective properties based on knowledge of the microstructure allows for a quantitative correlation between the changes in the microstructure and the macroscopic properties. When the complete details of the microstructure are not available, the effective properties are approximated using partial statistical information in the form of correlation functions yielding upper and lower bounds of the range in which the properties belong~\cite{Torquato1991}. Among the earliest bounds are the weighted arithmetic and harmonic means of the elastic moduli. These one-point bounds are known as the Voigt-Reuss bounds~\cite{Voigt1889573,Reuss192949,Pabst2015}. Voigt averaging assumes that all phases share the same strain, while Reuss averaging assumes that all phases have the same stress. Hashin and Shtrikman defined bounds that depend on two-point and higher-order correlation functions by using variational principles~\cite{Hashin1963127}. Hashin-Shtrikman bounds provide more precise limits on the effective elastic properties compared to Voigt-Reuss bounds. The Mori-Tanaka analytical method assumes a continuous matrix phase with dispersed inclusions, and uses Eshelby's solution to approximate the local stress and strain fields around the inclusions. Averaging these fields over the microstructure results in the effective elastic tensor~\cite{Mori1973571}.  Most of these analytical or numerical models assume simple geometry for second-phase particles such as sphere and ellipse ~\cite{Bakshi20112615, Mori1973571, Hashin1963127}. The simplified geometry for  irregular-shaped second phase particles gives the unpredictable result as reported by Bakshi \textit{et al.}~\cite{Bakshi20112615}. Attempts are made to derive properties through image processing of microstructure ~\cite{Gokhale19992369, Garboczi19991, Torquato2000411} but the models are based on mean-field and other spatial averaging techniques. The mean-field model is hardly predictive, especially in the case of statistical extremes. One way to overcome this difficulty is to do direct computation using all the available microstructure information through micrographs. Numerical methods like FEA provide reliable predictions of the effective elastic tensor by modeling the geometry and properties of the constituent phases. FEA is particularly suited for analyzing materials with complex microstructures~\cite{LI2020100769}. It can accurately represent intricate geometries, irregularly shaped phases, interpenetrating phases, and overlapping inclusions. 
 

In general, the microstructure of a material is a complex ensemble of different phases which exist in all sizes and shapes and have different orientations and crystallography. Predicting macroscopic behaviour directly from microstructure while considering all the complexity is a challenging task. Object Oriented Finite Element Method (OOFEM)~\cite{Langer200115} is an appropriate tool for such purpose as it effectively incorporates the information of various phases.
Kumara \textit{et al.}~\cite{Kumara2018529} modelled the elastic properties of nickel-based alloy with OOFEM and showed that it could capture the anisotropy in Young’s modulus. Gupta \textit{et al.}~\cite{Gupta201524} studied the mechanical behaviour of Mg-Li alloy with OOFEM and established its utility for multiscale modelling. Effective elastic modulus for $Ni-Al_{2}O_{3}$ composite~\cite{Sharma2014320} and turtle shell~\cite{Balani20111440} has also been reported with OOF$2$. OOF$2$ also has its utility in residual stress ~\cite{Chawla2002395,Vedula20012947, Molaro2015255,Zimmermann19993155}, and crack propagation ~\cite{Chawla2002395, Vedula20012947, Yousef20052809} studies. The scope of OOF$2$ is not limited to structural mechanics. It can also be used to determine thermal conductivity as reported by Bakshi \textit{et al.} for Al-Si-CNT composite ~\cite{Bakshi2010419}.   

This paper is organized as follows. The formalism to implement Phase-field method, extracting the Free energy from CALPHAD, and Finite element method for elastic moduli calculation of the microstructure is presented in Section ~\ref{model}. Section~\ref{numerical} provides the details of numerical simulations. Section~\ref{result} contains the results and discussions. First, we elaborate on microstructures obtained by the phase-field method in Section~\ref{PFM}. This is followed by Finite element analysis in Section~\ref{FEM_Results}.  Finally, Section~\ref{conclusion} concludes the article.  

\section{Model Formulation}
\label{model}
The following sections describe the modelling methods used in the present work. 

\subsection{Phase-field method}
\label{calphad}
We employ one conserved order parameter ($c$), and three non-conserved order parameters ($\eta_1$, $\eta_2$ and $\eta_3$) to describe the Ni-Al microstructure. The composition($c$) distinguishes between the $\gamma$ and $\gamma'$ phases. The ordered $\gamma'$ phase, however, has four structural variants. These four ordering types are energetically equivalent to each other. The three non-conserved order parameters contribute to describing the type of L1$_2$ variant in the $\gamma'$ phase.

To understand the state of the system as a function of time, we require two kinds of dynamic equations which govern the evolution of these order parameters. For conserved order parameter composition ($c$), the time evolution equation is the modified diffusion equation, written as follows, 

\begin{align}
  \frac{\partial c}{\partial t} &= \nabla .[M\nabla \mu_{total}],
  \label{1}
 \end{align}
 where, $M$ denotes the mobility, $c$ represents the (scaled) composition, $t$ designates time and the total chemical potential is expressed as $\mu_{total}$ = $\mu_{chem}$ + $\mu_{elast}$.
 Assuming isotropic diffusivity by using scalar mobility,
 eqn.~\ref{1} becomes,
 \begin{align}
  \frac{\partial c}{\partial t} &= M\nabla^2 \mu_{total} + \zeta_{\phi_c},
  \label{NI1}
 \end{align}
 where, $\zeta_{\phi_c}$ is the Langevin noise term. The expression for $\mu_{total}$ is given by  
 \begin{align}
  \mu_{total} = \frac{\delta [\frac{F_{total}}{N_V}]}{\delta c} = \frac{\delta [\frac{F_{chem} + F_{elast}}{N_V}]}{\delta c},
 \end{align}
 where $(\delta/\delta c)$ is a variational derivative, $N_V$ refers to the atoms present in a unit volume. The evolution equation for each non-conserved order parameter is expressed as,
 \begin{align}
  \frac{\partial \eta_i}{\partial t} &= -L \frac{\delta F_{total}}{\delta \eta_i} + \zeta_{\eta_i}, i=1,2,3, 
  \label{NI2}
 \end{align}
where, $\zeta_{\phi_i}$ is the Langevin noise term. 
The overall free energy consists of chemical and elastic contributions. These are separately discussed in sections~\ref{chemical} and ~\ref{elastic}.
 
\subsubsection{Chemical contribution to the Chemical potential}
\label{chemical}
 The chemical part of the free energy has bulk and gradient contributions. The bulk free energy $f(c,\eta_1,\eta_2,\eta_3)$ representative of Ni-Al system is approximated by,
 \begin{align}
 f(c,\eta_1,\eta_2,\eta_3) = 
 &{ \frac{A_1}{2}(c-c_1)^2 + \frac{A_2}{6}(c_2-c) (\eta_1^2+\eta_2^2+\eta_3^2)} \nonumber \\   
 &{ + \frac{A_3}{3} (\eta_1\eta_2\eta_3) + \frac{A_4}{24} (\eta_1^4 + \eta_2^4
+\eta_3^4) } \nonumber \\
 &{ + \frac{A_5}{24} (\eta_1^2\eta_2^2 + \eta_2^2\eta_3^2 +\eta_1^2\eta_3^2 ) } , 
 \label{Eqn1a}
 \end{align} 
where, $(\eta_1,\eta_2,\eta_3) = (-1,-1,1),(1,-1,-1),(-1,1,-1),(1,1,1)$ correspond to four free energy minima~\cite{Vaithyanathan20024061,WANG19982983}. These sets satisfy the following condition, 
\begin{equation}
 \frac{\partial f(c,\eta_1,\eta_2,\eta_3)}{\partial \eta_i} = 0, \text{where }  i = 1,2,3.
 \label{Eqn2a}
\end{equation}

The process of finding parameters A$_1$ - A$_5$ of eqn.~\ref{Eqn1a} for performing a quantitative study is discussed in section ~\ref{FE_basics}. 

\subsubsection{Elastic contribution to the Chemical potential}
\label{elastic}
Assuming Hooke's law is valid for both phases, we can describe the elastic free energy as 
\begin{eqnarray}
\label{eq:FElast}
F_{elast} &=& \frac{1}{2}\int_{V}\sigma_{ij}^{el}(r)\varepsilon_{ij}^{el}(r) dV \nonumber
\\&=& \frac{1}{2}\int_{V}\varepsilon_{ij}^{el}(r) C_{ijkl}(r)\varepsilon_{kl}^{el}(r) dV,
\end{eqnarray}
where $C_{ijkl}$ denotes the elastic stiffness tensor. The elastic strain $\varepsilon_{ij}^{el}$ is calculated  by subtracting the eigen strain $\varepsilon_{ij}^{0}$ from the total strain  $\varepsilon_{ij}$, 
\begin{equation}
  \label{eq:ElasticStrain}
\varepsilon_{ij}^{el} = \varepsilon_{ij} - \varepsilon_{ij}^{0}.
  \end{equation} 
The total strain $\varepsilon_{ij}$ is computed by 
\begin{equation}
\label{eq:TotalStrain}
\varepsilon_{ij} = \frac{1}{2}\left\lbrace \frac{\partial \vec u_{i}(\vec r)}{\partial r_{j}} 
+ \frac{\partial \vec u_{j}(\vec r)}{\partial r_{i}} \right\rbrace ,
\end{equation} 
where $\vec u(\vec r)$ denotes the displacement field. \textcolor{black}{The composition-dependent eigen strain can be computed as
\begin{equation}
\label{eq:EigenStrain}
\varepsilon_{ij}^{0}(c) = \beta(c)\varepsilon^{T}\delta_{ij} ,
\end{equation}
where $\beta(c)$, given by $c^3(10-15c+6c^2)$, interpolates smoothly between 0 and 1. $\varepsilon^{T}$ is the magnitude of eigen strain and $\delta_{ij}$ denotes the Kronecker delta. \textcolor{black}{The stiffness $C_{ijkl}(c)$ is expressed as
\begin{equation}
\label{eq:Stiffness}
C_{ijkl}(c) = C_{ijkl}^{0} + \alpha(c)\Delta C_{ijkl} ,
\end{equation}
where $C_{ijkl}^{0}$ is defined as $\frac{1}{2}(C_{ijkl}^{\alpha_2}+C_{ijkl}^{\alpha_1})$, $\alpha(c)$, given by $c^3(10-15c+6c^2)-\frac{1}{2}$, interpolates smoothly for the two phases. $\Delta C_{ijkl}$ is given by the difference in the phases' elastic moduli, i.e., $\Delta C_{ijkl} = C_{ijkl}^{\alpha_2} -C_{ijkl}^{\alpha_1}$. In the case of homogeneous elasticity, $\Delta C_{ijkl}$ = 0.
}
The following equation ensures mechanical equilibrium at all times,
  \begin{equation}
  \label{eq:MechEq}
  \nabla \ . \ \sigma_{ij}^{el} = \frac{\partial \sigma_{ij}^{el}} {\partial r_{j}} = 0 .
  \end{equation}
We employ an iterative method for solving the mechanical equilibrium equation (Eq.~\ref{eq:MechEq})~\cite{Gururajan20075015}. First, the displacement field is computed assuming $C_{ijkl} (c) = C_{ijkl}^{0} $.} The zeroth order solution is further used to solve the inhomogeneous elasticity problem. A non-zero value of $\Delta C_{ijkl}$ exists after the zeroth step. The displacement field is used to subsequently compute the strain field and elastic energy ($F_{elast}$). From this, we obtain the elastic contribution to the chemical potential ($\mu_{elast}$). 

\subsection{Deriving Free energy from CALPHAD}
\label{FE_basics}
A phase-field model relies on free energy to define a system's framework. It is essential to consider correct near-real values for the free energy parameters. 
The free energy $f(c,\eta_1,\eta_2,\eta_3)$ representing the system under study is given by Eqn. \ref{Eqn1a}. If we consider a single order domain i.e. where $\eta_1 = \eta_2 = \eta_3 = \eta$, then Eqn. \ref{Eqn2a} becomes, 


\begin{equation}
 f(c,\eta) = \frac{b_0}{2}(c-c_1)^2 + \frac{b_2}{2}(c_2-c) \eta^2 + \frac{b_3}{3} \eta^3 + \frac{b_4}{4} \eta^4.
 \label{Eqn3}
\end{equation}
The above equation is a qualitative representation of the free energy of our system. To extend this equation for a quantitative study, we compare the free energy curves for both the phases, obtained from CALPHAD with the above equation to find the coefficients $b_0,b_2,b_3,b_4,c_1 \text{and }c_2$. The barrier height $\Delta f$ at $c=c^*$ is equal to 1. To adjust all parameters w.r.t. this reference, we divide the parameters $b_0$, $b_2$, $b_3$ and $b_4$ by $f(c^*,0)$ . The new values correctly represent the Ni-Al system at different temperatures. The parameters obtained after fitting are tabulated in Table~\ref{Table1}.



\begin{table}[ht]
\begin{center}
\caption[Estimated Free Energy parameters at different temperatures]{Estimated parameters at different temperatures  }
\begin{tabular}{c c c c} \hline \hline
Parameter & 900 K & 1000 K & 1100 K  \\ 
\hline

$c_{\gamma}$  & 0.1125  & 0.1210  & 0.1300  \\
$c_{\gamma'}$ & 0.2245  & 0.2231  & 0.2216   \\
 \hline

$c_1$     & 0.1125  & 0.1210  & 0.1300    \\
$c_2$     & 0.3798  & 0.3764  & 0.3711    \\
$b_0$     & 970.31  & 1151.06  & 1336.12   \\
$b_2$     & 217.16  & 235.23  & 244.98     \\
$b_3$     & -174.10  & -179.54  & -176.52  \\
$b_4$     & 140.37  & 143.73  & 140.14    \\

 \hline     \hline \end{tabular}
                         
\label{Table1}
\end{center}
\end{table} 

\begin{figure}[h]
\centerline{\includegraphics[clip, trim=0cm 0cm 0cm 0cm, width = 0.9\columnwidth]{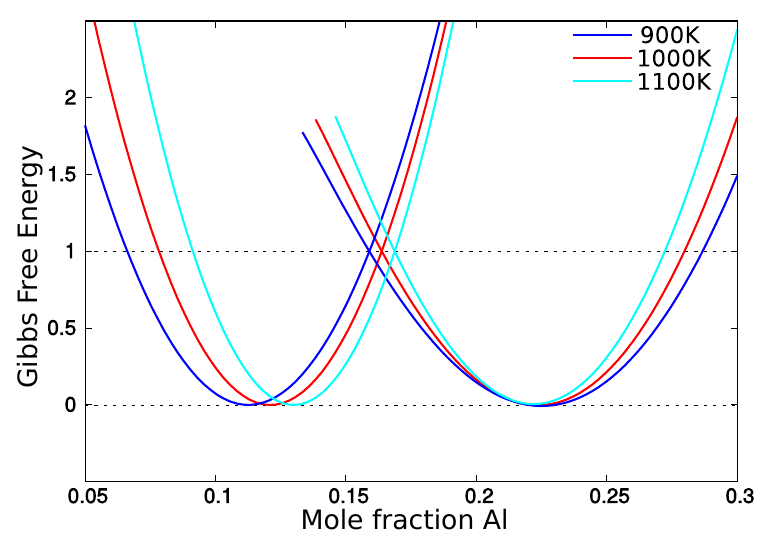}}
\caption[Gibbs Free energy versus composition curves for $\gamma$, and $\gamma'$ phases at 900 K, 1000 K and 1100 K]{Gibbs Free energy versus composition curves for $\gamma$, and $\gamma'$ phases at 900 K, 1000 K and 1100 K, obtained by substituting the calculated parameters listed in from Table~\ref{Table1}. } 
\label{Fig2_Mix}
\end{figure}

The free energy curves obtained after substituting these parameters in Eqn.~\ref{Eqn3} can be seen in Fig.~\ref{Fig2_Mix}. The equilibrium compositions $c_{\gamma}$ and $c_{\gamma'}$ are accurately represented for all three temperatures. 

\subsection{Finite element method}
\label{FEM}
\begin{figure}[htbp]
	\centering
		\includegraphics[width=0.95  \linewidth]{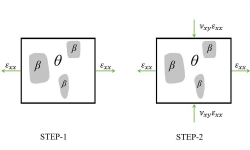}
		\caption{A schematic figure showing the application of the boundary conditions in two steps for the evaluation of effective elastic properties.}
	\label{fig:Schematic_boundary_conditions}
\end{figure}

The effective elastic properties of the alloys were evaluated by finite element analysis of the virtual micrographs generated via Phase-field Modelling.  The finite element analysis (FEA) was accomplished using an open-source OOFEM software, OOF$2$. For doing FEA of microstructure, we need a mesh which is a good approximation of the morphology of the  microstructure. OOF$2$ provides a good combination of image-processing, meshing, finite element solver and post-processing tools altogether. Consider a two-phase microstructure, as shown in Figure~\ref{fig:Schematic_boundary_conditions}, with the phases $\theta$ and $\beta$. First, the image file (of $m \times n$ pixels) is imported in OOF$2$, and the phases of the microstructure were identified using their pixel values and grouped under individual pixel groups. Next, the anisotropic elastic properties of the $\theta$ and $\beta$ phases were defined in terms of the components of the stiffness matrices and assigned to the respective pixel groups. It was also assumed that the local coordinate systems associated with the $\theta$ and $\beta$ phases were aligned perfectly with the global coordinate system. This was followed by building a 'skeleton' which defines the geometry of the mesh. Building a proper skeleton is a critical step in the finite element analysis, and determines how well the resulting mesh from the skeleton approximates the microstructure. For this purpose, OOF$2$ defines an effective energy functional (E) for the mesh~\cite{Langer200115}. The skeleton adapts to the morphology of the microstructure by minimizing the energy functional. The energy functional encompasses two types of fictitious energy, one is homogeneity energy $E_{homog}$ which takes account of the  material property in an element and another is shape energy $E_{shape}$ which takes account of the geometry of element, as given by~\cite{Langer200115}
 \begin{equation}
               E = \alpha E_{homog} + (1-\alpha) E_{shape}
               \label{euqation 0}
 \end{equation}

 \noindent The variable $\alpha$ shows the relative importance of shape and homogeneity energy. The effective energy functional takes value in the range of 0 to 1 for each element. The quality of an adapted skeleton is measured by the homogeneity index, which is a measure of the number of materials shared by an element, i.e. an element that has pixels of $n$ different phases material has a homogeneity index of $1/n$.  Since the microstructures were periodic, care was taken to make the mesh periodic as well. The field equations to be solved by OOF$2$ were chosen for a plane stress analysis. The boundary conditions to evaluate the effective elastic modulus (say, along the $x$ or the horizontal direction) associated with a microstructure were applied in two steps. In the first step, uniaxial displacements (in terms of pixels), simulating a tensile test (along $x$-axis), were applied to the right and to the left edges of the microstructure, while the top and the bottom edges were allowed to deform freely. The finite element model was solved using a static, iterative solver in OOF$2$.

The average normal elastic strains in the $x$-direction (${\overline{\epsilon}_{xx}}$) and the $y$-direction ($\overline{\epsilon}_{yy}$) were obtained as output, and the Poisson's ratio ($\nu_{xy}$) was evaluated using the equation,
\begin{equation}
\nu_{xy} = -\frac{\overline{\epsilon}_{yy}}{{\overline{\epsilon}_{xx}}}.
 \label{euqation 01}
\end{equation}
In the second step, a new set of boundary conditions are introduced in addition to the tensile deformation to the left and right edges, simulating compression for the top and the bottom edges. 
The Poisson's ratio calculated in the first step is now used to apply compressive displacements on the top and the bottom edges of the meshed microstructure, with the compressive displacement being equal to $\frac{\nu\ast \overline{\epsilon}_{xx}\ast n}{2}$ pixels. The application of the boundary conditions in two steps is demonstrated in Figure~\ref{fig:Schematic_boundary_conditions}. 
The effective elastic modulus (along $x$-direction) is evaluated by dividing the average normal stress in the $x$-direction by the average normal strain in the $x$-direction, using $\frac{\overline{\sigma}_{xx}}{\overline{\epsilon}_{xx}}$. A similar analysis was also performed to calculate the effective elastic modulus along $y$-direction.

\section{Simulation Parameters}
\label{numerical}

We represent the system on a simulation grid of $512\times512$ grid points. The simulation domain is periodic, and step size is equal to 4.0 in both $x$ and $y$ directions.  The timestep is taken as 0.1, and the mobility ($M$) and gradient energy coefficient are equal to 1.  \textcolor{black}{The $\gamma$-$\gamma'$ coherent interfacial energy is considered as 10 $mJ m^{-2}$~\cite{WANG19982983}}. The characteristic length ($L'$), energy ($E'$) and time ($T'$) are employed for obtaining the non-dimensionalized parameter values. 

For FEA, building the skeleton was initiated by having a grid size of $75$ triangular elements along the edge normal to the $x$-direction, and $75$ triangular elements along the edge normal to the $y$-direction. Next, all elements with heterogeneity less than $0.9$ were selected and refined by a process of bifurcation of the elements. During the process of refinement of the selected heterogeneous elements, equal importance was given to the shape of the elements and the heterogeneity of the elements by setting an energy parameter $\alpha$ to $0.5$. The process of refinement of the heterogeneous elements was repeated till the homogeneity index of the entire skeleton was above $0.96$. The resulting skeleton was checked for the presence of badly-shaped elements and was corrected. 

To check if the system size has any effect on the finite element analysis, images of the same microstructure but of different pixel dimensions, for example, $256 \times 256$, $512 \times 512$, and $1024 \times 1024$, were performed. The effect of composition on the effective elastic properties was evaluated by FEA of microstructures of Ni-Al alloy, generated for Aluminium content of \textcolor{black}{0.1618, 0.1721, and 0.1823 mole percent} obtained at a temperature of $1000$ K. The effect of temperature on the effective elastic properties was evaluated by FEA of microstructures obtained at a fixed Aluminium concentration of \textcolor{black}{0.1721 mole percent} for three temperatures, namely $900$ K, $1000$ K, and $1100$ K. The effect of aging time, which led to the growth of the $\gamma'$ phase, on the effective elastic properties was also investigated.

 
\section{Results and Discussion}
\label{result}  
This section discusses the results obtained from phase-field method and its FEM analysis.  
\subsection{Microstructures obtained from phase-field method}
\label{PFM}
We study the microstructure evolution under isothermal condition and carry out a temporal analysis of the entire process. 
\subsubsection{Effect of Composition}
\begin{figure}[!ht]
 \centering
  \includegraphics[clip, trim=0cm 0cm 0cm 0cm, width = 1.0\columnwidth]{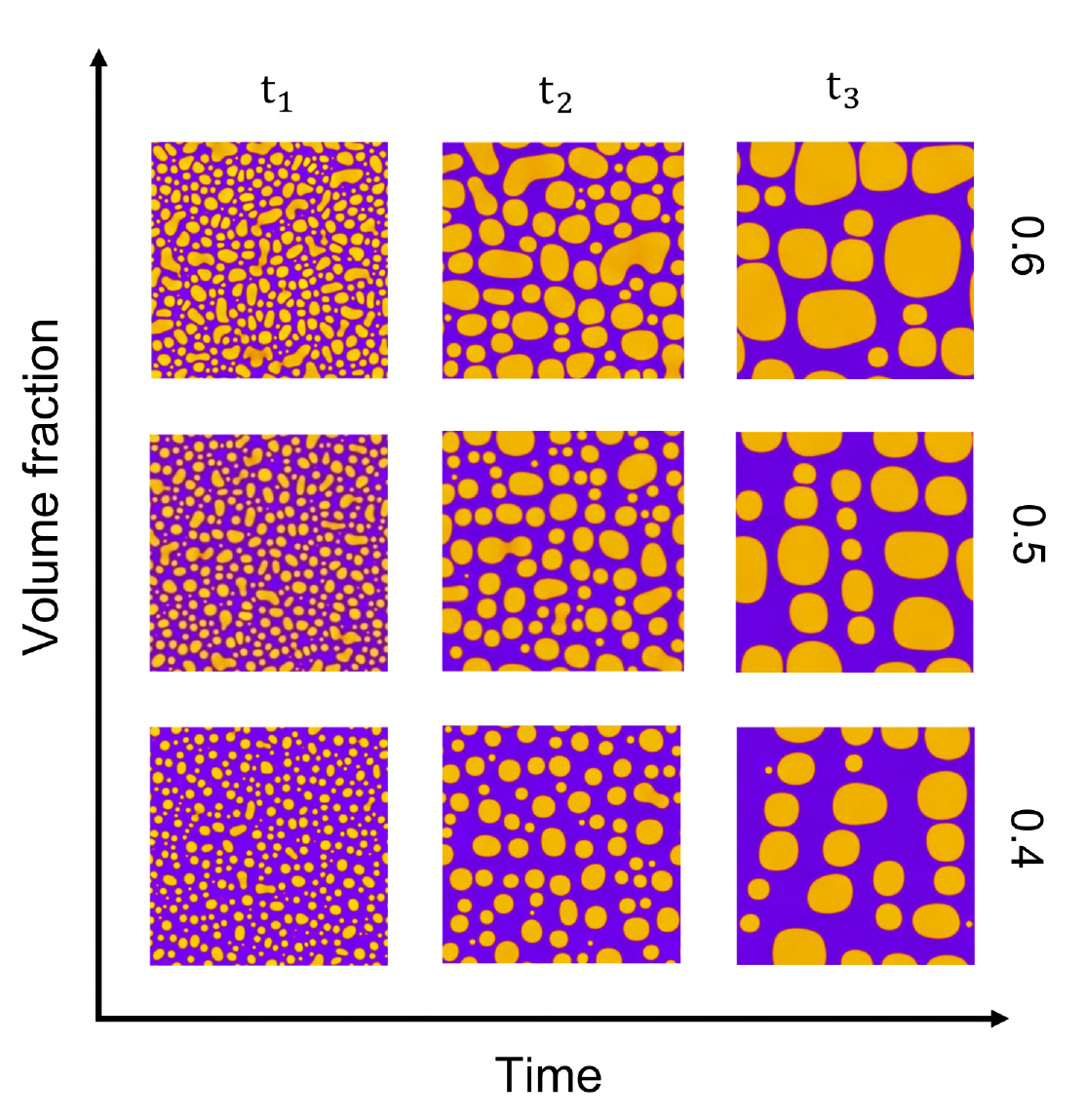}
  \caption{Variation map of simulated microstructures of the Ni-Al alloy with aging time for the equilibrium composition of 0.1618, 0.1721, and 0.1823 Al mole fraction corresponding to $\gamma'$ equilibrium volume fraction of 0.4, 0.5, and 0.6 respectively at 1000 K at initial, intermediate and final times during the simulation. }
  \label{B_2}
\end{figure} 

\begin{figure}[htbp]
 \centering
   \includegraphics[width=0.98 \linewidth]{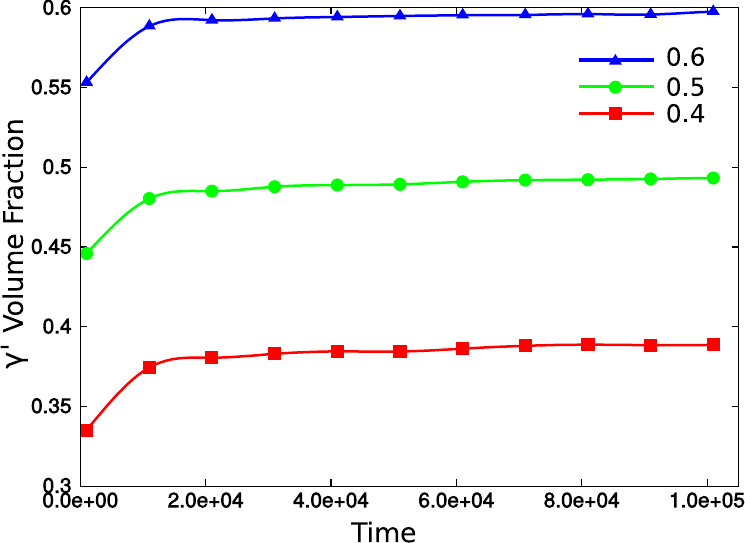}
   \caption{Variation of $\gamma'$ volume fraction with aging time for three  equilibrium $\gamma'$ volume fraction of 0.4, 0.5, and 0.6 respectively at 1000 K  showing growth and coarsening regions. }
   \label{B_4}
\end{figure} 
 Simulations are carried out for three different alloys of composition ($X_{Al}$) 0.1618, 0.1721, and 0.1823, corresponding to 0.4, 0.5 and 0.6 volume fractions of $\gamma'$ phase at 1000 K. The alloys are denoted by points A, O and B in Figure~\ref{NiAl_PD}.

At the onset, a conserved Langevin noise is introduced in the alloy, so the overall alloy composition remains unaffected. Similar noise terms are also introduced for the non-conserved order parameters $\eta_1$, $\eta_2$, and $\eta_3$. These fluctuations in $c$ and $\eta$ are responsible for the nucleation to take place. After the formation of sufficient nuclei, the sustained noise terms are removed~\cite{Vaithyanathan20024061}. It is evident that some smaller particles dissolve to make way for the growth of larger particles, as shown in Figure~\ref{B_2}. Particles adjacent to each other with dissimilar sets of non-conserved order parameters do not coalesce. This happens to avoid the formation of an anti-phase domain boundary with higher energy than two standard interfaces combined. Particles with identical variants can coalesce if they are close to each other. Simultaneously, larger particles coarsen at the cost of smaller ones.

Although the alloy compositions correspond to 0.4, 0.5, and 0.6 volume fractions of the $\gamma'$ phase according to the CALPHAD data, Figure~\ref{B_4} shows
that the volume fractions attained are slightly lower than those predicted by the lever rule. The Gibbs-Thompson effect is responsible for such a shift in composition with curvature further translating into a change in volume fraction. 
 
 \subsubsection{Effect of Temperature}
\begin{figure}[!ht]
\centering
  \includegraphics[clip, trim=0cm 0cm 0cm 0cm, width = 1.0\columnwidth]{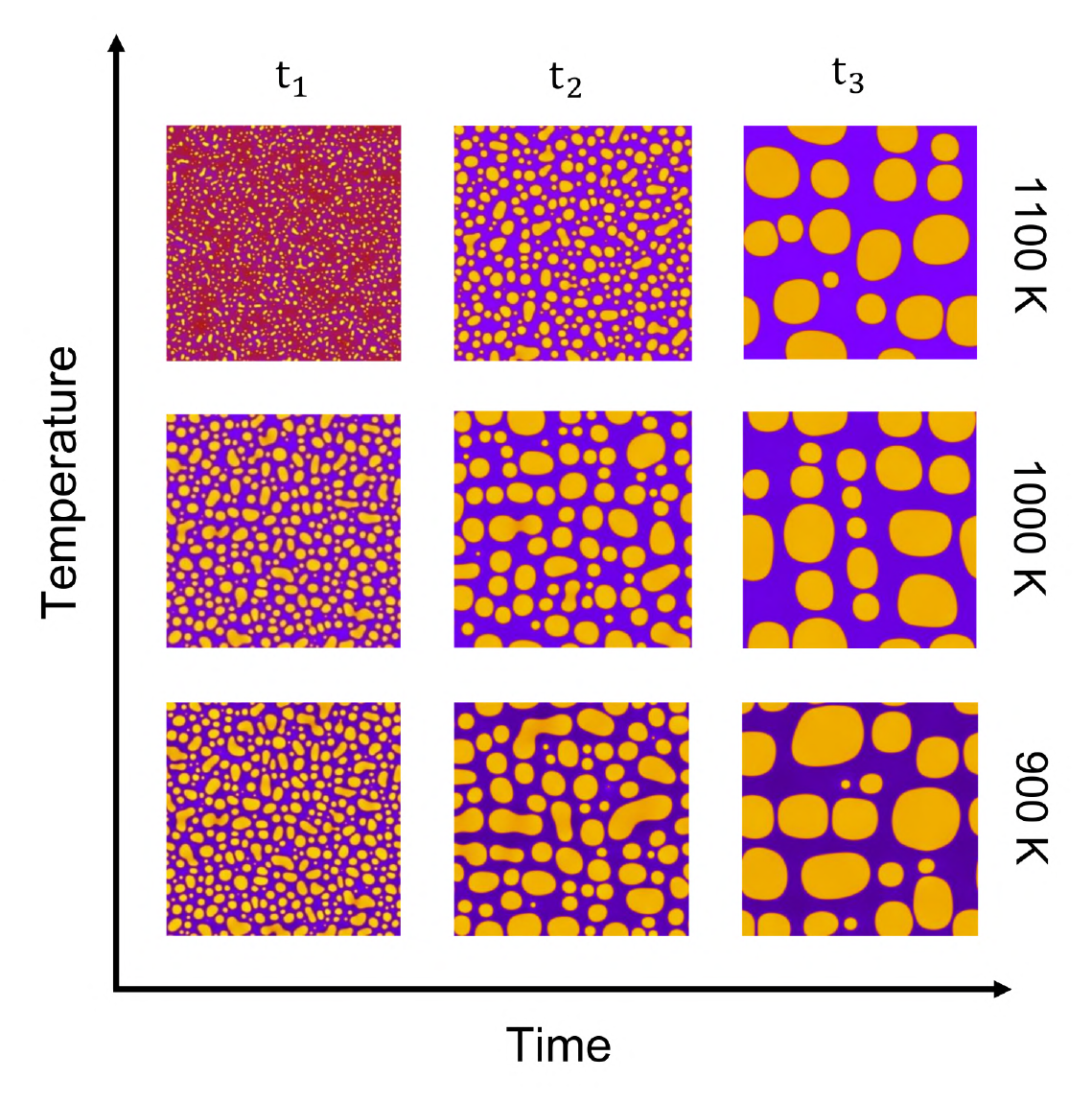}
 \caption{Variation map of simulated microstructures of the Ni-Al alloy with aging time for three isothermal temperatures of 900 K, 1000 K, and 1100 K  at an equilibrium composition of 0.1721 Al mole fraction corresponding to $\gamma'$ equilibrium volume fraction of 0.5 for the three aging times.}
 \label{B_1}
\end{figure} 

\begin{figure}[!ht]
 \centering
   \includegraphics[width=0.98 \linewidth]{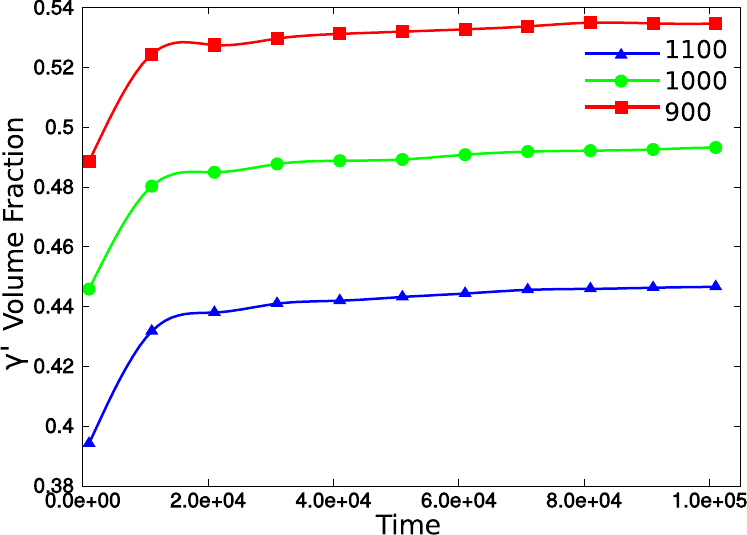}
   \caption{Variation of $\gamma'$ volume fraction with aging time for alloys C, O, and D corresponding to temperatures of 900 K, 1000 K, and 1100 K.}
   \label{B_3}
\end{figure} 

Isothermal simulations are carried out corresponding to three temperatures (900 K, 1000 K, 1100 K) with alloy composition fixed at 0.1721. Points C, O, and D in Figure~\ref{NiAl_PD} refer to the systems in the scope of our study. The temporal evolution of microstructure for these alloys is discussed in this section.

The initial condition for all three temperatures is a common composition. Thus, their composition maps would be identical at zero time. Even though the alloy composition is the same, the equilibrium compositions are different for the three temperatures. This generates a difference in volume fractions of $\gamma'$ phase at the three temperatures. Figure~\ref{B_1} shows the temporal evolution of microstructures. The time $t_1$ of Figure~\ref{B_1} shows the composition maps after nucleation. As discussed in the previous section, a sustained Langevin noise term aids the process of nucleation. Thus, nucleation and growth of super-critical nuclei take place. At $t_2$ of Figure~\ref{B_1}, we witness the mid regime of coarsening. The particles align, and coarsening begins. Few pairs of particles with identical variants coalesce. Others with different structural variants display coarsening. Late-stage coarsening can be observed at $t_3$. Alignment in horizontal rows and vertical columns is evident at this time step. It is interesting to see that in spite of the same starting composition, the microstructures evolve to be different. This is not only due to the difference in equilibrium compositions but also because of the difference in the elastic properties. It is observed from Figure~\ref{B_3} that the volume fraction of $\gamma'$ particles decreases as temperature increases from 900 K to 1100 K. This is due to the shifting of equilibrium compositions as the temperature is varied.

\subsection{Effective elastic properties from FEM }
\label{FEM_Results}

\begin{figure}[h]
	\centering
\includegraphics[width=0.9\linewidth]{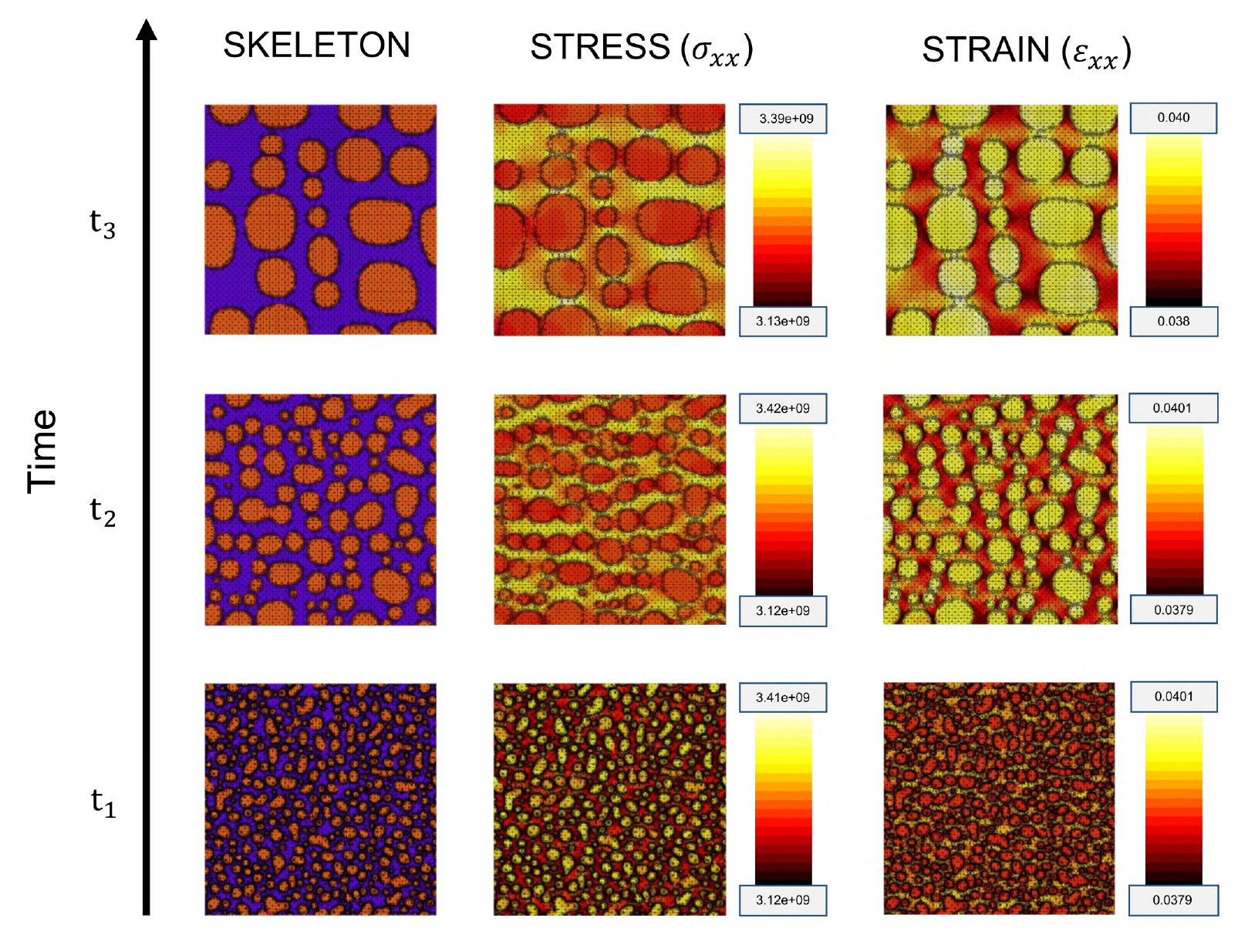}
\caption{Variation map of simulated microstructures of the Ni-Al alloy along with aging time for equilibrium $\gamma'$ 0.5 volume fraction at $1000$ K  with their respective adapted skeletons and distribution of $\sigma_{xx}$ and $\epsilon_{xx}$ obtained after applying uni-axial tensile displacements along the horizontal (or $x$) direction.}
	\label{fig:Stress_strain_distribution_0pt5_1000K}
\end{figure}

The effective elastic properties (elastic modulus and Poisson's ratio) of the nickel-aluminium alloy obtained from the FEA are presented in this section. Figure~\ref{fig:Stress_strain_distribution_0pt5_1000K} shows the various steps of finite element analysis using OOF$2$ for the microstructures obtained for three aging times at $1000$ K for the Ni-Al alloy with an equilibrium $\gamma'$ volume fraction of $0.5$. The adapted skeleton obtained during the pre-meshing step, and the distribution of $\sigma_{xx}$ and $\epsilon_{xx}$ obtained during finite-element post-processing are also shown for the respective microstructures. The effective properties are evaluated from the area-averaged stress ($\overline{\sigma}_{xx}$) and strains ($\overline{\epsilon}_{xx}$ and $\overline{\epsilon}_{yy}$) obtained from the stress and strain distributions. Similar analyses were performed for the microstructures obtained at $900$ K and $1100$ K for Ni-Al alloy with an equilibrium $\gamma'$ volume fraction of $0.5$, as well as for  equilibrium $\gamma'$ volume fractions of $0.4$ and $0.6$ at $1000$ K. These analyses are presented in S2, S3, S4, S5 
in the \textit{Supplementary Material}. Varying the initial mesh size didn't have a significant impact on the effective elastic properties; more information on mesh size is available 
in the \textit{Supplementary Material}.

\begin{figure*}[htbp]
\centering
\subfigure[]{\includegraphics[width=0.45  \linewidth]{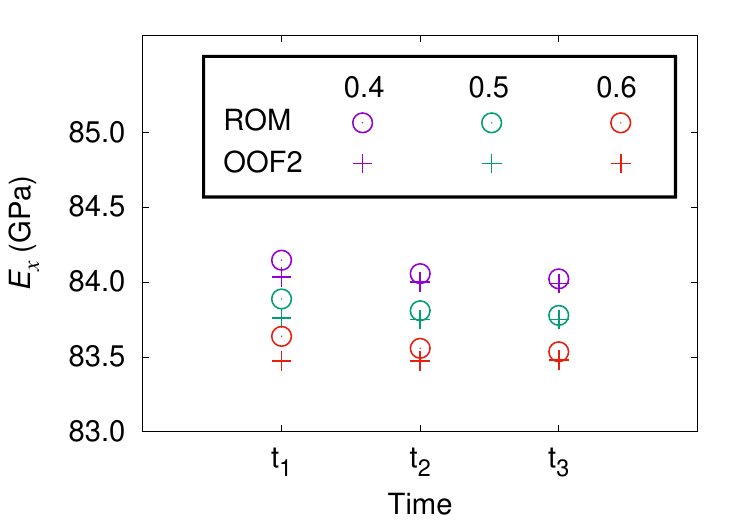}}
\subfigure[]{\includegraphics[width=0.45  \linewidth]{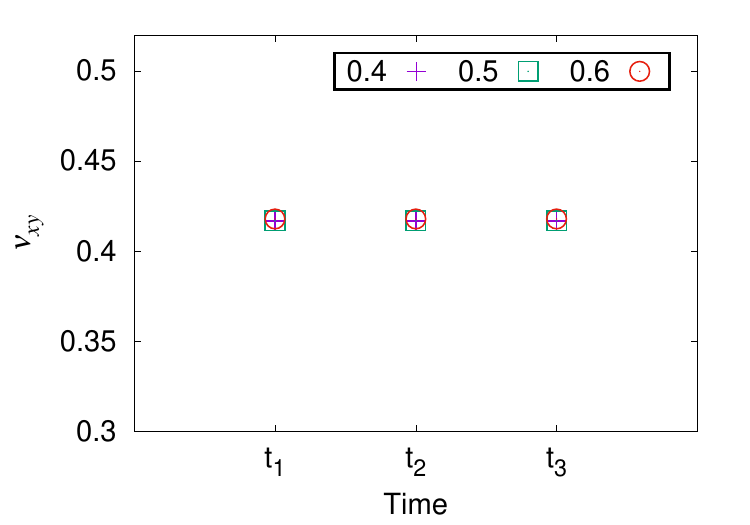}}
\caption{Variation map of effective elastic modulus and poison's ratio of the alloy along $x$-direction with aging times and for three equilibrium $\gamma'$ volume fraction at $1000$ K. Elastic moduli predicted by the rule of mixtures (ROM) are also shown for comparison.}
\label{fig:variation_map_Composition_1000K}
\end{figure*}

\begin{figure*}[h]
	\centering
	\subfigure[]{\includegraphics[width=0.45  \linewidth]{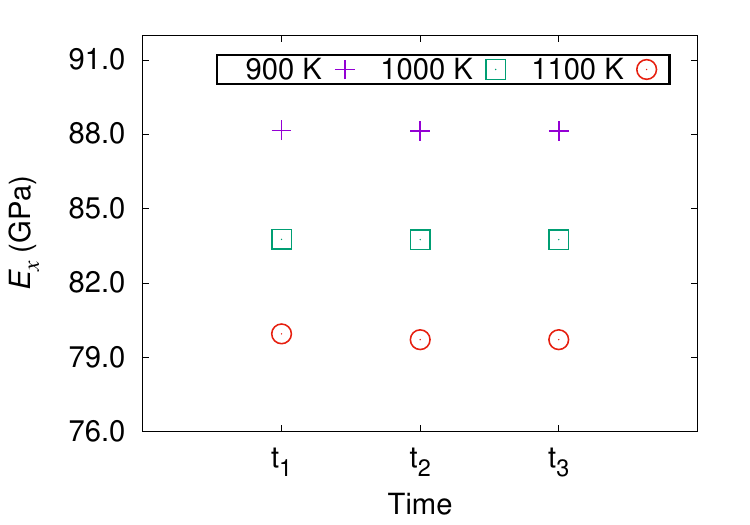}}
	\subfigure[]{\includegraphics[width=0.45  \linewidth]{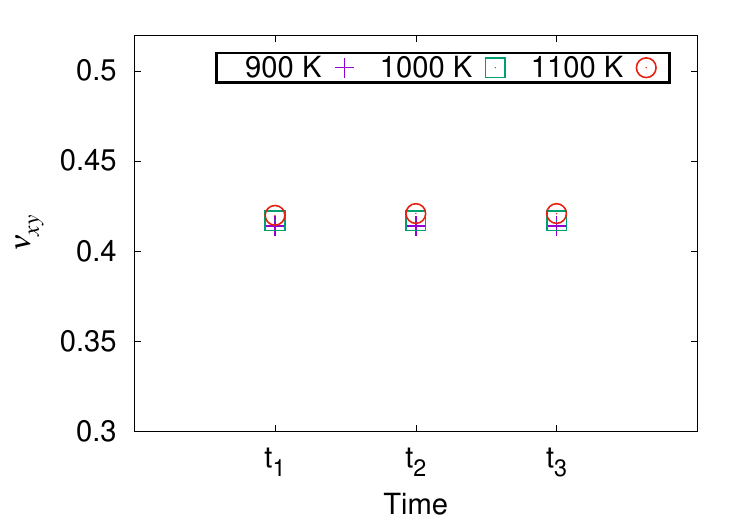}}		
	\caption{Variation map of effective elastic modulus and effective Poisson's ratio of the alloy along $x$-direction with aging times for three temperatures at  0.5 $\gamma'$ volume fraction.}
	\label{fig:variation_map_temperature_0.5}
\end{figure*}

\par
Figure~\ref{fig:variation_map_Composition_1000K} shows the variation map of  effective elastic modulus and effective Poisson's ratio of the alloy along $x$-direction with aging time for three equilibrium volume fractions of the $\gamma'$ phase, at $1000$ K. In Figure~\ref{fig:variation_map_Composition_1000K} (a),  in addition to the values of effective elastic modulus obtained from OOF2, values obtained using the simple rule of mixtures (ROM) are also presented to show that the values predicted by OOF2 remain well below the upper-limit of values of elastic moduli. The rule of mixtures can be expressed as 
\begin{equation}
{E_x}^{eff} = v_{\gamma}{E_x}^{\gamma} + v_{\gamma'}{E_x}^{\gamma'}.
\end{equation}
Here ${E_x}^{eff}$, ${E_x}^{\gamma}$ and ${E_x}^{\gamma'}$ refer to the elastic moduli along the $x$-direction for the alloy, the $\gamma$-phase, and the $\gamma'$-phase, respectively. $v_{\gamma}$ and $v_{\gamma'}$ are the volume fractions of the respective phases. The volume fractions of the $\gamma'$ phase refer to the points A, O, and B in the phase diagram shown in Figure~\ref{NiAl_PD}. Figure~\ref{fig:variation_map_Composition_1000K} shows that the elastic modulus of the alloy decreases with increasing volume fraction of the  $\gamma'$ phase at a fixed aging time. The effective elastic moduli of the alloys range from a maximum value of ~$84.03$ GPa for $\gamma'$ equilibrium volume fraction of $0.4$ at an aging time of $t_1$ to a minimum value of ~$83.47$ GPa for $\gamma'$ volume fraction of $0.6$ at an aging time of $t_1$ at $1000$ K. The decrease in the effective elastic modulus of the alloy with increasing $\gamma'$ content can be justified since the elastic modulus of the $\gamma'$ phase is ~$82.60$ GPa, which is lower than that of the $\gamma$ phase (with an elastic modulus of ~$84.93$ GPa) at $1000$ K. It is also observed that the volume fraction of $\gamma'$ increases with the increase of aging time at a given temperature, details of which are provided in \textit{Supplementary Material}. This explains the observed slight decrease in the effective elastic moduli of the alloy as the aging time is increased. The variation map of effective elastic modulus along the $y$-direction (${E_y}^{eff}$) with aging time for three equilibrium volume fractions of the $\gamma'$ phase, at $1000$ K also shows similar values and trends as that of ${E_x}^{eff}$, as shown in Figure S10(a) 
 in the \textit{Supplementary Material}.
\par
 In Figure~\ref{fig:variation_map_Composition_1000K}(b) it is observed that, unlike the effective elastic moduli, the effective Poisson's ratio does not show any variation with aging time and equilibrium composition and remains constant at a value of ~$0.42$. The  variation map of effective Poisson's ratio, $\nu_{yx}$ with aging time for three equilibrium volume fraction of the $\gamma'$ phase, at $1000$ K also shows similar trend and values as that of $\nu_{xy}^{eff}$, and the plots are provided in Figure S10(b) 
 in the \textit{Supplementary Material}. 

Figure~\ref{fig:variation_map_temperature_0.5} shows the variation map of effective elastic modulus and effective Poisson's ratio of the alloy along $x$-direction with aging times for three temperatures at $\gamma'$ equilibrium volume fraction of $0.5$. Figure~\ref{fig:variation_map_temperature_0.5} (a) indicates that while the effective elastic moduli (${E_x}^{eff}$) decrease with increasing temperature, varying the aging time does not lead to a significant difference in properties at a given temperature. The effective Poisson's ratio ($\nu_{xy}^{eff}$) does not show any significant variation with changing aging time and temperature. ${E_y}^{eff}$ and $\nu_{yx}^{eff}$ also show similar trends and values as ${E_x}^{eff}$ and $\nu_{xy}^{eff}$, respectively, when plotted with aging time for the three temperatures. These Figure S11(a) and S11(b)
are available in the \textit{Supplementary Material}.

The plane stress and plane strain conditions form bounds for the effective properties. Since in our case, we assume a bulk material, a plane stress condition is considered. For the sake of completion, we have also calculated effective elastic properties for all 15 alloys in our simulation space with plane strain conditions. These results are summarized in Figure S12 and S13 
in the \textit{Supplementary Material}

\section{Conclusions}
\begin{itemize}
    \item In the present study, we successfully implement a physics-based multiscale modelling method for the Ni-Al system. Our model transfers information between different length scales by integrating CALPHAD with a quantitative Phase-field model over which an image processing-based Finite Element Method (FEM) is applied to predict mechanical behaviour directly from simulated microstructure.
    
    \item With the change in aging temperature from 900 K to 1100 K for an alloy of known composition, the volume fraction of $\gamma'$ phase decreases. It is in agreement with the phase diagram determined using the CALPHAD approach. The phase-field model considers the change in equilibrium compositions due to the Gibbs-Thompson phenomenon, which eventually drives the coarsening of $\gamma'$ phase.
    
     \item The effective elastic modulus of Ni-Al alloy predicted from FEM calculation is a vital function of aging temperature and the volume fraction of $\gamma'$ phase. Coarsening does not have a significant impact on effective elastic modulus.
    
    \item Effective Poisson's ratio shows meager variation with aging temperature and no dependence on coarsening or change in volume fraction of $\gamma'$ phase.

    \item \textcolor{black}{The predicted properties clearly establish structure-processing-(elastic) property relationships for the Ni-Al alloy considered in this work, and can be utilized for understanding the elastic response of components made from Ni-based superalloy under harsh operating conditions.}
\label{conclusion}
    
\end{itemize}

\section*{Acknowledgements}
We acknowledge National Supercomputing Mission (NSM) for providing computing resources of ‘PARAM Sanganak’ at IIT Kanpur, which is implemented by C-DAC and supported by the Ministry of Electronics and Information Technology (MeitY) and Department of Science and Technology (DST), Government of India. 
RM is thankful for financial support received from Center for
Development of Advanced Computing (C-DAC) Project No. Meity/
R\&D/HPC/2(1)/2014. The authors are also grateful to the Computer Centre, IIT Kanpur for HPC facilities.


\section*{Data availability}
The raw/processed data required to reproduce these findings cannot be shared at this time as the data also forms part of an ongoing study.
\bibliography{mybibfile}

\end{document}